\documentclass[aip,reprint]{revtex4-1}

\usepackage{color}
\usepackage{graphicx}
\usepackage{amsmath}
\usepackage{subcaption}

\usepackage{natbib}

\begin{document}
\title{Magnetized Ablative Rayleigh-Taylor Instability in 3-D}

\author{C. A. Walsh}
\email{walsh34@llnl.gov}
\affiliation{Lawrence Livermore National Laboratory}

\date{\today}

	\begin{abstract}
		3-D extended-MHD simulations of the magnetized ablative Rayleigh-Taylor instability are presented for the first time. Previous 2-D simulations claiming perturbation suppression by magnetic tension are shown to be misleading, as they do not include the most unstable dimension. For perturbation modes along the applied field direction, the magnetic field simultaneously reduces ablative stabilization and adds magnetic tension stabilization; the stabilizing term is found to dominate for applied fields $>5T$, with both effects increasing in importance at short wavelengths. For modes perpendicular to the applied field, magnetic tension does not directly stabilize the perturbation, but can result in moderately slower growth due to the perturbation appearing 2-D (albeit in a different orientation to 2-D ICF simulations). In cases where thermal ablative stabilization is dominant the applied field increases the peak bubble-spike height. Resistive diffusion is shown to be important for short wavelengths and long timescales, reducing the effectiveness of tension stabilization.
	\end{abstract}
	\maketitle
	
	\section{Introduction}

	\begin{figure}
		\centering
		\includegraphics[width=0.5\textwidth]{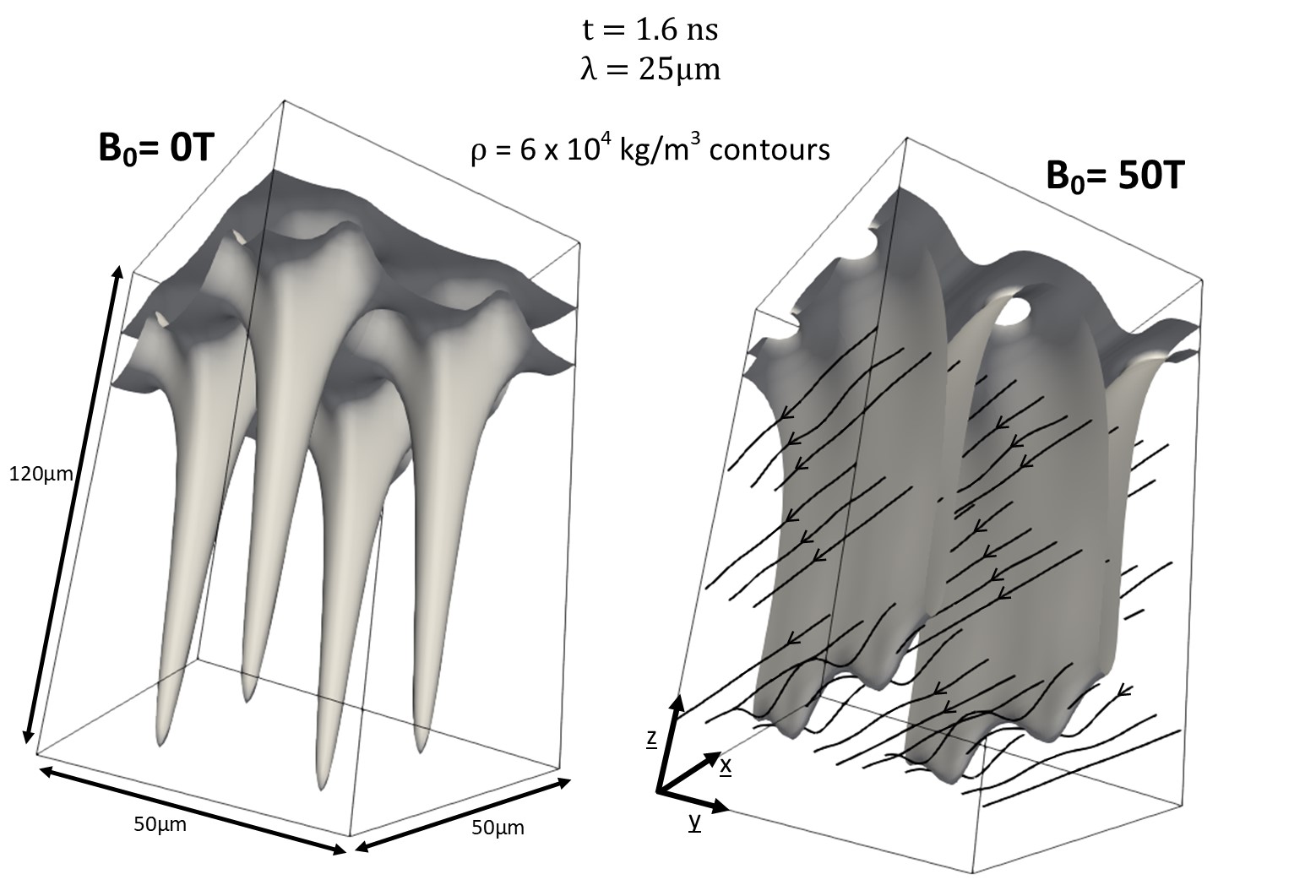}
		\caption{ \label{fig:single_contours} Density contours ($6 \times 10^{4} kg/m^3$) at 1.6ns for simulations with $25\mu m$ perturbation wavelengths. On the left is without an applied field and on the right is using a $50T$ magnetic field along $\underline{x}$. The perturbation is almost completely stabilized by magnetic tension in $\underline{x}$ but remains unsuppressed in $\underline{y}$. 2-D simulations ignore the $\underline{y}$ direction.}
	\end{figure}
	
	Magnetic fields can be applied to inertial confinement fusion (ICF) implosions to reduce hot-spot energy losses by magnetizing the electron population. This was demonstrated for direct-drive implosions at the OMEGA Laser Facility, where an 8T applied field increased the yield and temperature of a spherical capsule by 30\% and 15\% respectively \cite{chang2011}. An externally-applied field has also been predicted to bring current best-performing implosions on The National Ignition Facility into the burning plasma regime \cite{perkins2013,perkins2017}. MagLIF experiments on the Z-facility are designed specifically with the use of an external magnetic field, allowing for slower and higher-adiabat designs \cite{gomez2020}; 2-D simulations predict yield amplifications by magnetization of $>100$ when sufficient laser preheat energy is used \cite{slutz2018}. In parallel, a campaign using a scaled-down version of MagLIF is underway on OMEGA \cite{barnak2017}. It is also thought that an applied magnetic field can assist fast ignition designs, as the field allows for enhanced coupling of charged particle beams to the hot-spot \cite{Wang2015}
	
	While the main focus of research to date has been on the effect of magnetic fields on the bulk hot-spot performance, it is important to also consider how magnetizing the plasma can change deceleration-phase perturbation growth. 2-D simulations have predicted the reduction of Rayleigh-Taylor due to magnetic tension \cite{srinivasan2013,perkins2017,PhysRevE.104.L023201}. As magnetic tension preferentially stabilizes short wavelength perturbations \cite{chandrasekhar1962} it has been suggested that mix could be reduced by the application of a magnetic field \cite{srinivasan2013}. Other simulations have demonstrated that an applied field reduces ablation of perturbations and can therefore also decrease stability \cite{walsh2019}.
	
	This paper presents the first 3-D magnetized ablative Rayleigh-Taylor (magARTI) instability simulations, demonstrating that a 2-D approach always ignores the least stable dimension. Magnetic tension cannot completely stabilize the Rayleigh-Taylor instability, as the field only suppresses modes where $\underline{k} \cdot \underline{B} \ne 0$. Instead, a large magnetic field is found to, at best, only moderately reduce perturbation growth ($\approx 1/3$ when thermal conduction is less important) and, at worst, significantly enhance short wavelength perturbation growth. This result, among others, has significant implications for the design of magnetized ICF implosions, where 2-D simulations are routinely employed (e.g OMEGA spherical implosions \cite{davies2015}, OMEGA cylindrical  \cite{davies2017,walsh2021exploring}, NIF spherical \cite{perkins2013,perkins2017}, MagLIF cylindrical \cite{slutz2010,slutz2012,slutz2016,slutz2018} and fast ignition spherical \cite{johzaki2016}).
	
	The 3-D nature of the Rayleigh-Taylor instability with an embedded magnetic field has been studied previously in astrophysical plasmas \cite{stone2007}; similar to the enclosed work, the magnetic field is found to stabilize the instability only along field lines, resulting in a striated unstable interface. The distinction of the work presented here is the applicability to ICF implosions, where ablation of dense material is key to tamping the instability growth.

	%\note{don't bother with full relation. Just state the x and y components. Introduce them as we go along, will make it more readable. Also state geometry, i.e. acceleration in z}
	%The linear magARTI growth rate can be written \cite{betti2001,chandrasekhar1962,masse2007}:
	
	%\begin{equation}
%\gamma = \sqrt{\frac{\alpha g}{\lambda + L_m} - \frac{\delta (\underline{B}\cdot\underline{\hat{k}})^2}{\lambda^2 \mu_0 (\rho_l+\rho_h)}} -  \beta\frac{ (\sqrt{D}+1) V_{a,B=0}}{\lambda}\frac{\kappa_{\perp}}{\kappa_{\parallel}} \label{eq:gamma}
%\end{equation}
	
%	Where $g$ is the acceleration, $\lambda$ the perturbation wavelength, $\underline{\hat{k}}$ the normalized direction under consideration, $L_m$ the density length-scale, $\underline{B}$ the magnetic field and $\alpha$, $\beta$, $\gamma$ are constants that depend on the geometry. The ablation velocity $V_{a,B=0}$ is the ablation velocity without thermal conduction suppression, i.e. $V_a = V_{a,B=0} \kappa_{\perp}/\kappa_{\parallel}$ \cite{betti2001}. Note this doesn't account for an increased hot-spot temperature which is often associated with the application of magnetic fields \cite{chang2011,walsh2019}. The factor $D$ represents the anisotropy of the thermal conductivity at the unstable interface \cite{masse2007}:
	
%	\begin{equation}
%		D = \frac{\kappa_{\parallel} \underline{\hat{b}}\cdot\underline{\hat{k}}+ \kappa_{\perp} (1-\underline{\hat{b}}\cdot\underline{\hat{k}})}{\kappa_{\perp}}
%	\end{equation}
	
%	Where $\underline{\hat{b}}$ is the magnetic field unit vector.
	The classical Rayleigh-Taylor instability, where a low density fluid is accelerated into a higher density fluid, is maximally unstable at short perturbation wavelengths \cite{zhou2017,zhou2019}. However, energy transport from the light to the dense fluid can induce mass ablation, which stabilizes the Rayleigh-Taylor growth \cite{sanz1994}, particularly for short perturbation wavelengths; this is then called the ablative Rayleigh-Taylor instability (ARTI), where the highest modes are classically unstable but also the most stabilized by ablation. In an ICF hot-spot this ablation manifests primarily through electron thermal conduction from the hot-spot core to the dense assembled fuel, although radiation transport and $\alpha$-heating also contribute \cite{Tong_2019}. In the linear regime, a perturbation is stable if the ablation velocity $V_{abl} > \sqrt{g \lambda}/2\pi R$, although this simple relation ignores the impact of geometry and finite density scale length effects \cite{betti2001}. When the ablation is dominated by thermal conduction, $V_{abl}$ is proportional to the thermal conductivity \cite{betti2001}. This can be shown simply by quantifying the volume of cold plasma heated up to the core hot-spot temperature every second ($\Delta T n_e V_{abl} \delta S$) by a heat-flow through a surface ($\delta S \kappa \nabla T_e$). This yields:
	
	\begin{equation}
		V_{abl} \sim \frac{\kappa^c T^{5/2}}{R \rho} \label{eq:Vabl}
	\end{equation}
	
	where $\kappa^c$ is the non-dimensional thermal conductivity that is independent of temperature and density and $R$ is the hot-spot radius. Overall this shows that the ablative stabilization is strongest for high temperature plasmas and perturbations with short wavelengths \cite{modica2013}. 
	
	Magnetized heat-flow in an MHD plasma follows \cite{braginskii1965,sadler2021}:
	
	\begin{equation}
		\underline{q} = -\kappa_{\parallel} \left(\underline{\hat{b}} \cdot \nabla T_e \right) \underline{\hat{b}} -\kappa_{\bot} \underline{\hat{b}} \times \left(\nabla T_e \times \underline{\hat{b}}\right)  - \kappa_{\wedge} \underline{\hat{b}} \times \nabla T_e\label{eq:heatflow}
	\end{equation}
	where $\underline{\hat{b}}$ is the magnetic field unit vector. The first term ($\kappa_{\parallel}$) represents unrestricted heat-flow along magnetic field lines. The second term ($\kappa_{\bot}$) is heat-flow perpendicular to the field and is suppressed by magnetization. The Hall Parameter ($\omega_e \tau_e$) is a metric for the degree of magnetization, a product of the electron gyrofrequency ($\omega_e$) and electron-ion collision time ($\tau_e$). If a magnetic field is applied along the ARTI spike propagation axis then the thermal conductivity in equation \ref{eq:Vabl} is $\kappa_{\parallel}^c$ and is unaltered by the electron magnetization. If, however, the magnetic field is applied perpendicular to the spike propagation axis, then the thermal conductivity in equation \ref{eq:Vabl} is $\kappa_{\bot}^c$ and the ablation velocity is suppressed by electron magnetization, decreasing the ablative stabilization.
	
	The final term ($\kappa_{\wedge}$) in equation \ref{eq:heatflow} is Righi-Leduc heat-flow, which acts perpendicular to both the magnetic field and the temperature gradient. This term is largest for moderate magnetizations $\omega_e \tau_e \approx 1$. The effect of this component is less clear than the parallel or perpendicular components, and does not affect the overall ablation velocity. Instead the Righi-Leduc term is found to break the ablation velocity symmetry, resulting in spikes not propagating directly along the acceleration axis. 
	
		\begin{figure}
		\centering
		\begin{subfigure}[b]{0.5\textwidth}
			\centering
			\includegraphics[width=1.\textwidth]{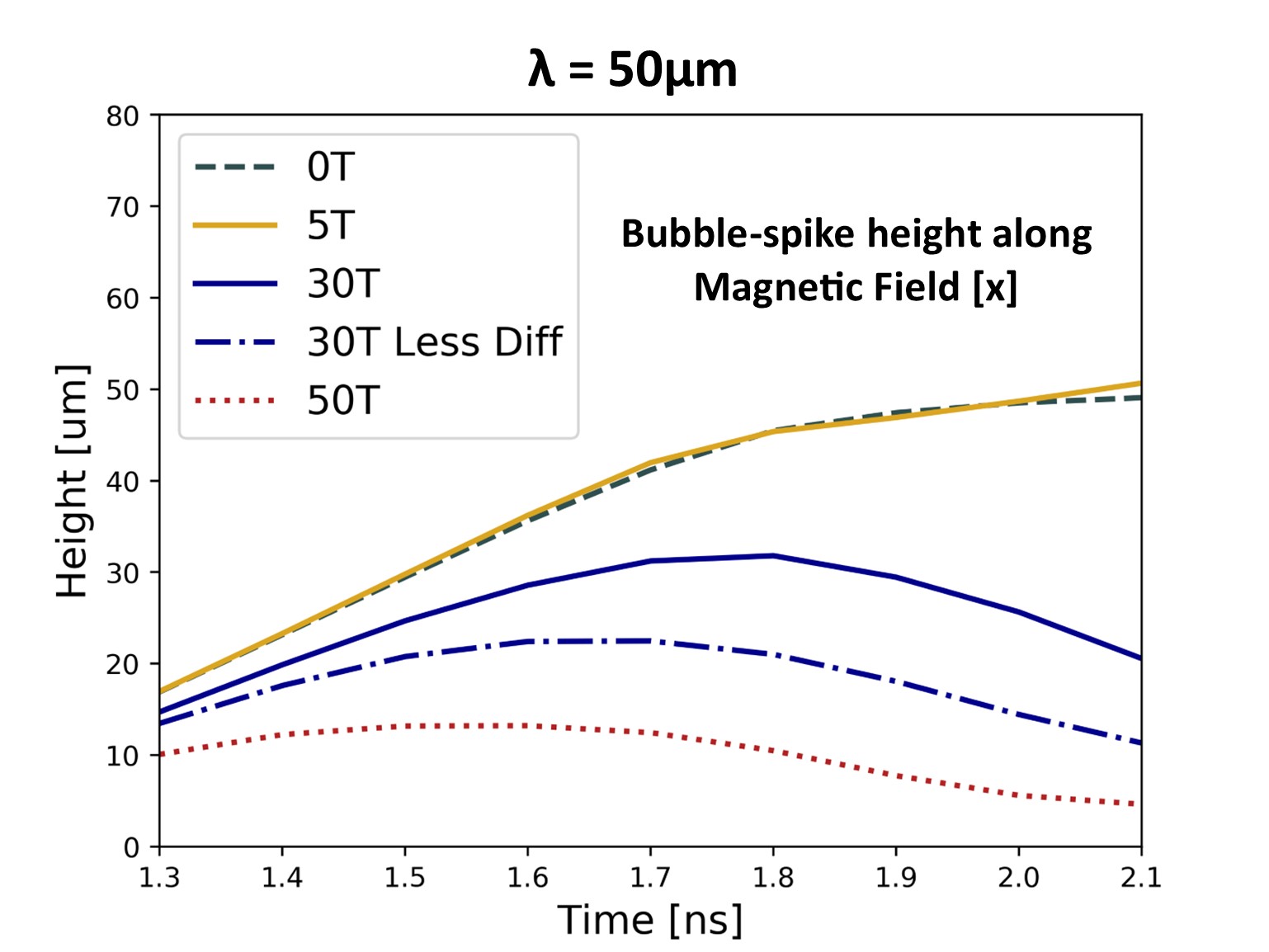}
			\caption{}
			\label{fig:single_temporalx}
		\end{subfigure}
		\begin{subfigure}[b]{0.5\textwidth}
			\centering
			\includegraphics[width=1.\textwidth]{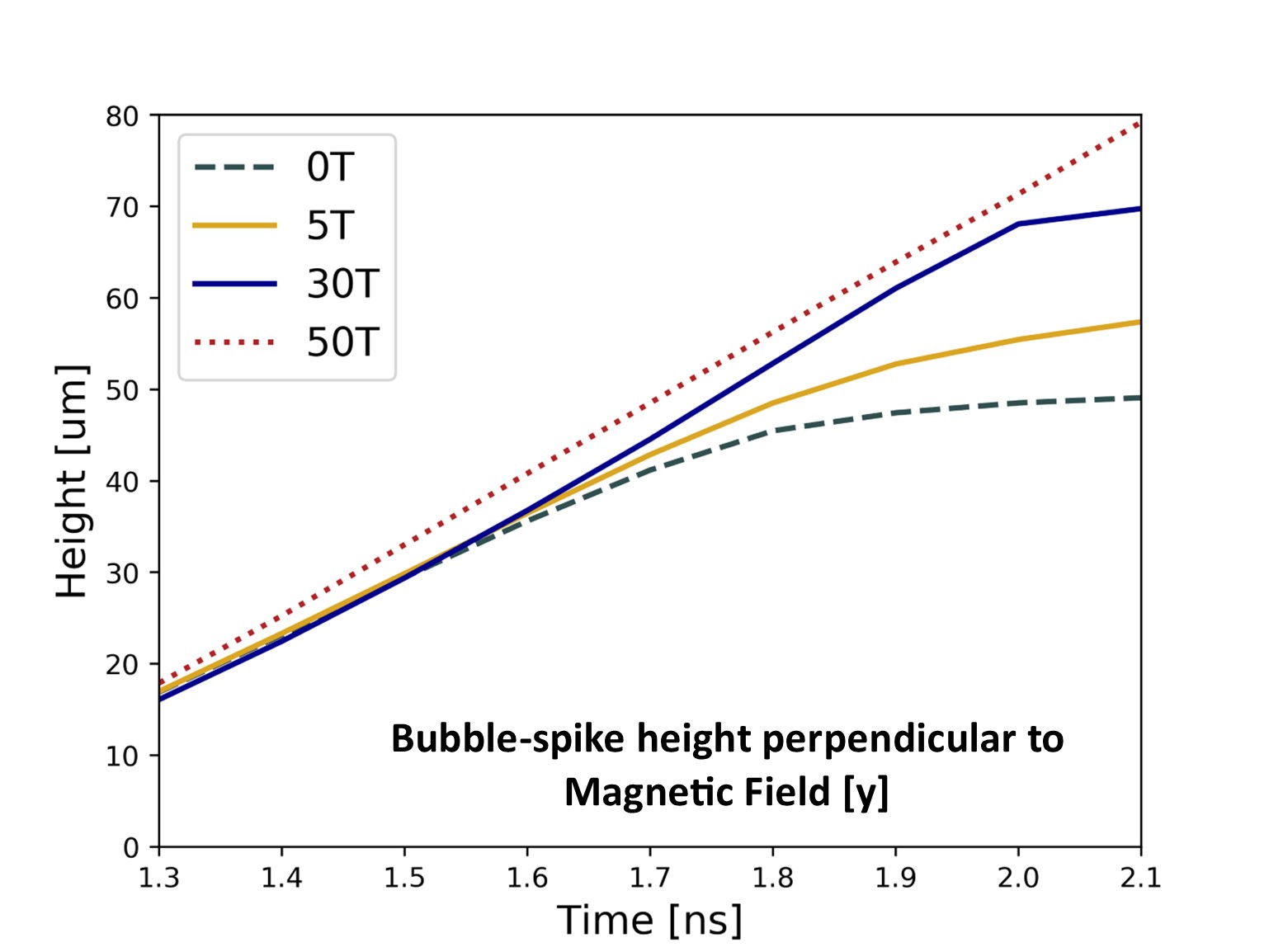}
			\caption{}
			\label{fig:single_temporaly}
		\end{subfigure}
		
		\caption{Bubble-spike height measured by the $1keV$ electron temperature contour as a function of time for a range of applied field strengths. The perturbation wavelength here is $\lambda = 50 \mu m$. a: bubble-spike height along the direction of the applied field ($\underline{x}$). Magnetic tension stabilizes the perturbation in this orientation. b: bubble-spike height perpendicular to the applied field ($\underline{y}$). Magnetization of the electrons results in reduced ablative stabilization at higher field strengths.}
		\label{fig:single_temporal}
	\end{figure}
	
	A magnetic field embedded in the plasma can also introduce a dynamically important Lorentz force ($\underline{j} \times \underline{B}$ where $\underline{j}$ is the current density). Typically the importance of the Lorentz force is characterized by the plasma $\beta$, which is the ratio of the thermal plasma pressure to the magnetic pressure. However, this can be misleading, as it assumes that the thermal pressure and the magnetic potential vary over the same scale length \cite{walsh2019}. The Lorentz force can be split into two components: a magnetic pressure ($\nabla  \underline{B} \cdot \underline{B}/2\mu_0$) and a magnetic tension ($\underline{B}\cdot\nabla \underline{B} / \mu_0$). This latter term acts to straighten the magnetic field lines and is strongest over short wavelengths \cite{chandrasekhar1962,srinivasan2013}. While high magnetic pressures in the hot-spot can reduce compression \cite{hansen2020,walsh2021exploring}, the magnetic tension is typically greatest around perturbations and can reduce vorticity \cite{walsh2019}.
	
	The stabilizing effect of the magnetic tension can be shown through simple energetic arguments. In the unmagnetized RT-instability, the growth of a dense spike to height $h$ results in the transfer of gravitational potential energy $(\rho_h-\rho_l)gh$ into kinetic energy, where $\rho_h$ and $\rho_l$ are the densities of the heavy and light fluids respectively. In a magnetized RT-instability, the work done on the magnetic field scales as $h | \underline{B} |^2 /\mu_0 \lambda$. Therefore, a magnetic field will suppress growth of perturbations with wavelengths: 
	\begin{equation}
		\lambda_{suppressed} <  \frac{| \underline{B} |^2}{\mu_0 (\rho_h-\rho_l)g}  \label{eq:lambda}
	\end{equation}
	This can be shown more thoroughly in the linear regime \cite{chandrasekhar1962}, although the scaling here suggests applicability in the non-linear regime as well. Suppression of instabilities along magnetic field lines has been demonstrated experimentally using gas-puff Z-pinches \cite{PhysRevE.104.L023201,lavine2021,conti2020}. However, the magnetic tension only stabilizes modes parallel with applied field direction, as perpendicular modes do not bend the field lines.

	%Magnetic fields embedded within a plasma result in the electron population taking curved trajectories between collisions. The degree of magnetization is characterized by the Hall Parameter $\omega_e \tau_e$, where $\omega_e$ is the gyrofrequency and $\tau_e$ is the electron-ion collision time. For $\omega_e \tau_e = 1$ the electrons take one full orbit on average between collisions; for $\omega_e \tau_e \gg 1$ the electrons are essentially confined to bulk motion along the magnetic field lines, greatly reducing the thermal energy transferred perpendicular to the magnetic field. \note{we should reduce this bit. too basic}
	
	The work presented here focuses on non-linear perturbations propagating perpendicular to the magnetic field orientation, which is applicable to the capsule waist of spherical implosions and everywhere for cylindrical designs with an axial magnetic field. The axis of spike growth is designated $\underline{z}$, while the magnetic field is oriented in $\underline{x}$. In this orientation magnetic tension acts against spike growth and also suppresses thermal conduction into the spike. While the thermal conduction is anisotropic (i.e. there is unrestricted heat-flow in $\underline{x}$ but not in $\underline{y}$), this particular effect is not found to induce anisotropic perturbation growth. The magnetic tension, on the other hand, is found to suppress perturbations with short wavelengths in $\underline{x}$ but not in $\underline{y}$.
	
	Another orientation relevant to the poles of magnetized spherical implosions is a magnetic field parallel to spike growth; previous simulations have shown that an increased hot-spot temperature can result in enhanced heat-flow at the spherical poles, resulting in lower perturbation growth \cite{walsh2019}. However, in this configuration the field tension does not significantly modify Rayleigh-Taylor, as the spike does not perturb the background field. For this reason, the work in this paper focuses on the case where the magnetic field is applied perpendicular to spike growth.
	
	The work here also focuses on perturbation growth during the stagnation-phase; applied fields are also expected to modify direct-drive perturbation growth during the drive-phase through magnetization of the laser conduction zone \cite{walsh2020a}. In indirect-drive, where radiation transport dominates over thermal conduction, an applied magnetic field does not modify the ablation process.

	The 3-D extended-MHD code Gorgon \cite{ciardi2007,chittenden2004,walsh2017} is used for this study. The simulations include magnetic transport by bulk advection, Nernst, cross-gradient-Nernst and resistive diffusion \cite{walsh2020}. Magnetic fields generated by the Biermann Battery mechanism and from ionization gradients\cite{sadler2020} are included; Gorgon has been used extensively to study self-generated magnetic fields in the deceleration phase of ICF implosions \cite{walsh2017,walsh2021}.  The Biermann Battery term in Gorgon has been benchmarked against magnetic fields observed in laser-foil interaction experiments \cite{campbell2021measuring,PhysRevLett.125.145001}. Improvements to the Epperlein \& Haines \cite{epperlein1986} transport coefficients are used \cite{sadler2021}, which have been shown to reduce the effect of magnetic field twisting in scenarios relevant to this paper \cite{walsh2021updated}. The magnetic transport in Gorgon has been successfully compared with cylindrical implosions probed by high energy protons \cite{knauer2010}, with particular sensitivity to bulk plasma advection and the Nernst term \cite{walsh2021exploring}. Thermal transport is treated anisotropically using a centered-symmetric scheme \cite{sharma2007}, which has been modified to include Righi-Leduc heat-flow; this scheme reduces erroneous diffusion of heat across the magnetic field lines \cite{sharma2007,walsh2018a}. The Lorentz force is included, as well as Ohmic dissipation of magnetic energy. 
	
	The 3-D magARTI test problem uses the constant acceleration $g=10^{15} m/s^2$ of a planar low density deuterium plasma (initial density $\rho_0 = 10^3kg/m^3$) into a dense deuterium plasma ($\rho_0 = 10^4kg/m^3$)).  At $t=0$ a $d=200\mu m$ transition layer between the two densities is used such that the density follows an exponential decay. The temperature of the low density plasma is initialized at $T_0 = 200eV$, with the high density plasma temperature set isobarically. Periodic boundary conditions are employed on these Cartesian simulations. The density gradient is set along $+\underline{z}$, while the acceleration is directed in $-\underline{z}$.
	
		\begin{figure}
		\centering
		\begin{subfigure}[b]{0.5\textwidth}
			\centering
			\includegraphics[width=1.\textwidth]{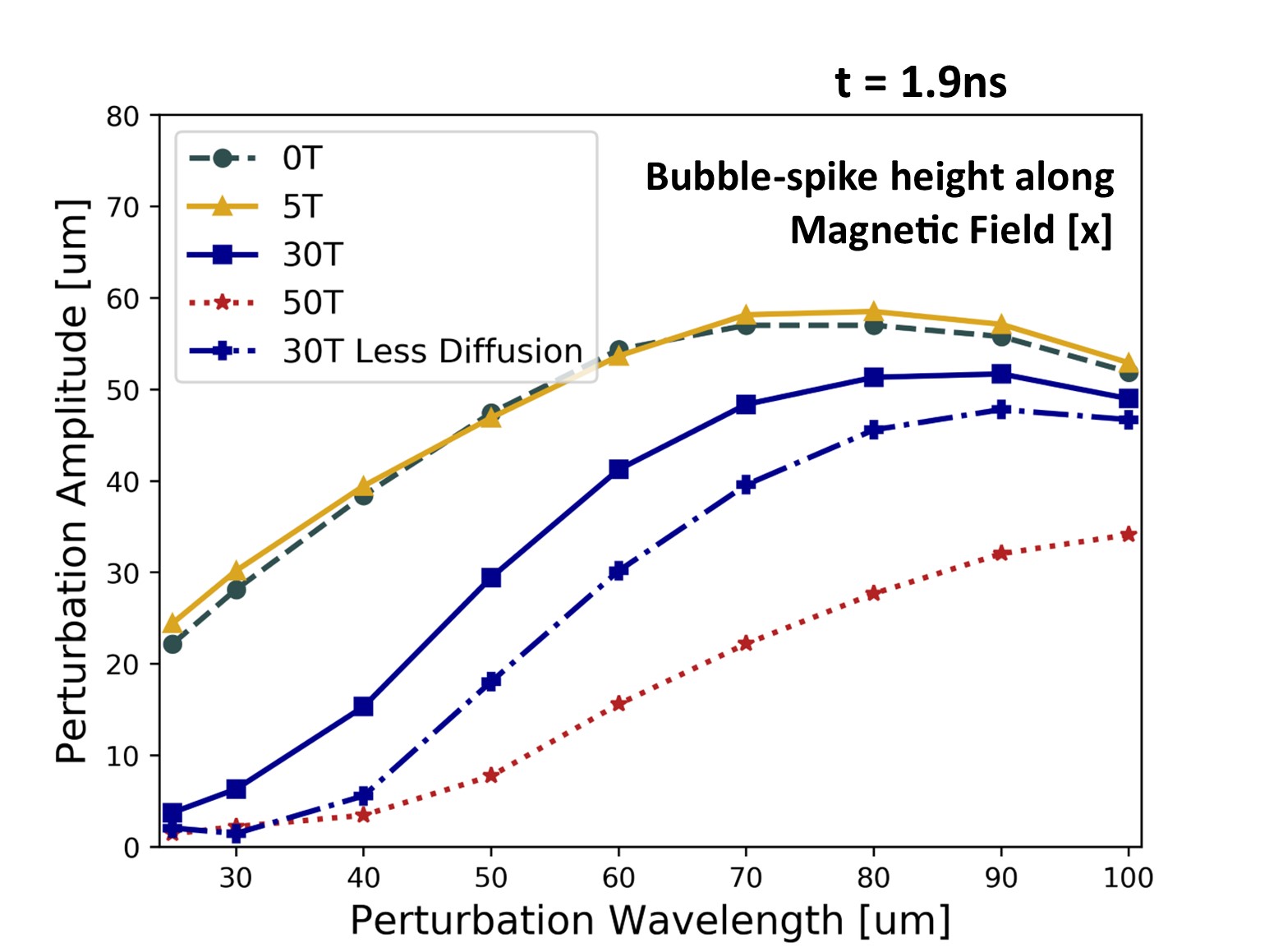}
			\caption{}
			\label{fig:single_wavelengthx}
		\end{subfigure}
		\begin{subfigure}[b]{0.5\textwidth}
			\centering
			\includegraphics[width=1.\textwidth]{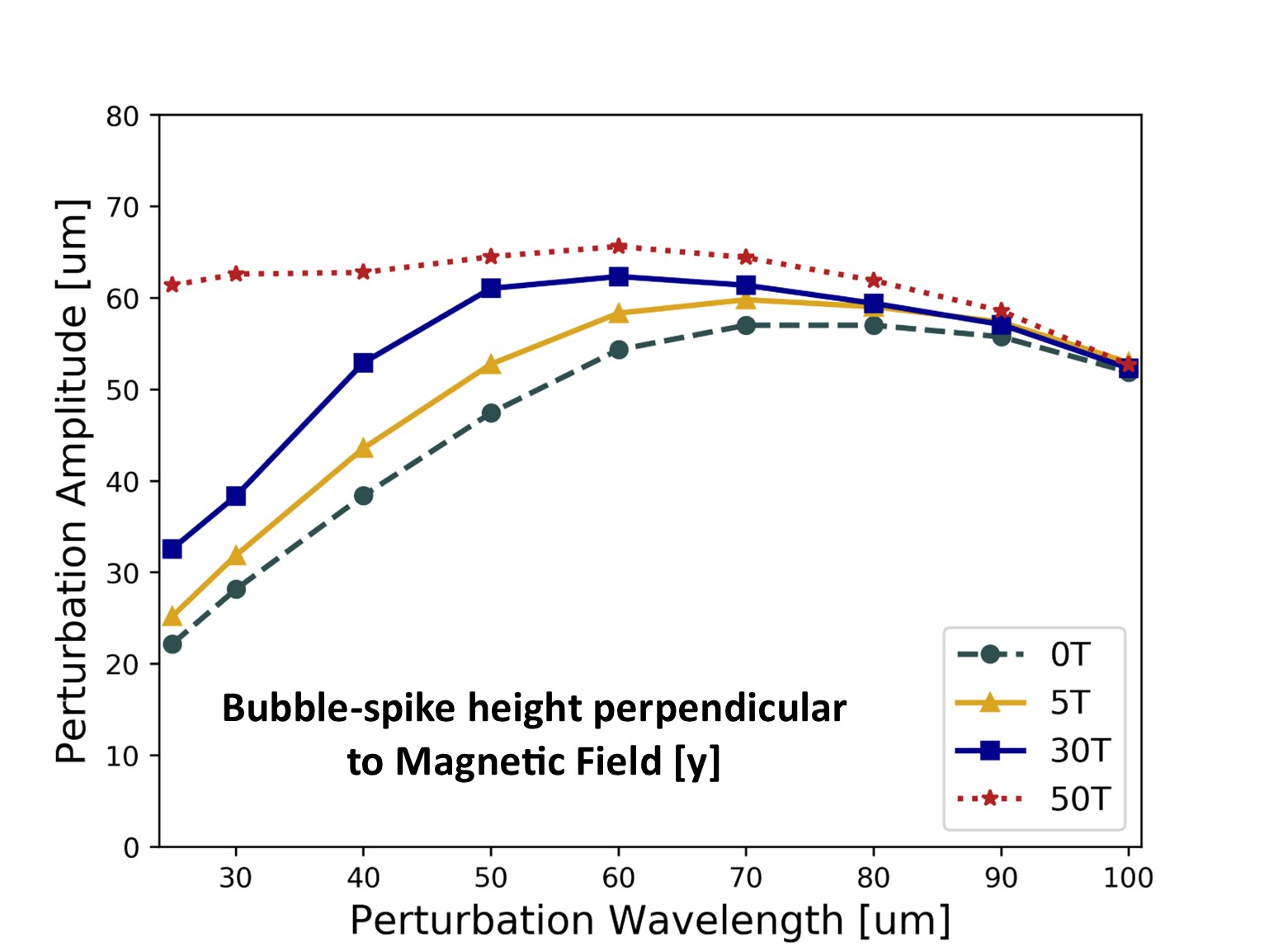}
			\caption{}
			\label{fig:single_wavelengthy}
		\end{subfigure}
		
		\caption{Single-mode bubble-spike height measured by the $1keV$ electron temperature contour for a range of perturbation wavelengths and applied fields. a: bubble-spike height along the direction of the applied field ($\underline{x}$). A 30T case where the magnetic resistivity has been artificially reduced by a factor of 10 is included to demonstrate the importance of this process in systems with large magnetic tension. b: bubble-spike height perpendicular to the applied field ($\underline{y}$). Ablative stabilization has increased importance at short wavelengths.  }
		\label{fig:single_wavelength}
	\end{figure}
	
	This setup inputs ($g$, $\rho_0$, $\Delta \rho_0$, $T_0$, $d$) were tuned to recreate key parameters of a decelerating hot-spot. By $t=1.6ns$ the peak density and temperatures are $10^5 kg/m^3$ and $4100eV$, with $\approx 100\mu m$ distance between the two peaks. At $t=1.9ns$ the peak density and temperature increase to $1.4 \times 10^5 kg/m^3$ and $5700eV$.
	
	A perturbation is applied by offsetting the transition layer between the dense and light plasma regions. For the single-mode perturbations this is an amplitude of $5 \mu m$, while the multi-mode calculations use $500$ randomly placed perturbations with random wavelengths and random amplitudes with a maximum of $\epsilon$ each, where $\epsilon$ is varied between $0.2\mu m$ and $1.0\mu m$ to assess the dependence of the observed phenomena to perturbation size.
	
	The magnetic field is applied along $\underline{x}$ using a strength proportional to the plasma density. This is in accordance with how the magnetic field would be compressed in planar or cylindrical geometry if the field is frozen into the plasma. The quoted applied magnetic field $B_0$ is the field strength that began in unheated ice  $|\underline{B}| =  B_0 \rho / \rho_{ice}$, where $\rho_{ice} = 250kg/m^3$. For example, $B_0 = 5T$ actually corresponds to an initialized field of $20T$ in the low density plasma and $200T$ in the high density plasma. For a spherical implosion the initial field will scale more weakly with density ($|\underline{B}| =  B_0 (\rho / \rho_{ice})^{2/3}$), although the results are found to be qualitatively similar. 

	Section \ref{sec:singlemode} first demonstrates the 3-D nature of the magnetized-ablative-Rayleigh-Taylor instability. The height of the spike along the magnetic field ($h_x$) is shown to be much lower than the height perpendicular to the magnetic field ($h_y$) when the magnetic tension is significant. Both temporal and wavelength dependence are explored in detail, showing the greatest perturbation anisotropy ($h_y/h_x$) for short wavelength perturbations. Resistive diffusion is also shown to play an important role in spike height along the field lines ($h_x$). Section \ref{sec:multimode} then demonstrates striated growth with multi-mode perturbations applied. An applied magnetic field is shown to reduce peak spike-bubble heights (i.e. both $h_x$ and $h_y$) for smaller initialized perturbations, analogous to moderate reductions in RTI growth rates in 2-D systems relative to 3-D.

	\section{Single-Mode \label{sec:singlemode}}

	Single-mode perturbations are applied using a $5\mu m$ amplitude cosine function, resulting in a spike down the simulation axis. Figure \ref{fig:single_contours} shows $\rho = 6 \times 10^4 kg/m^3$ density contours for $B_0 = 0T$ and $B_0 = 50T$ at 1.6ns for a perturbation wavelength of $25 \mu m$. In the magnetized case some magnetic field streamlines have been plotted to demonstrate the magnetic field bending. For $50T$ the spike is almost completely suppressed when looking along the $\underline{x}$ direction, which is consistent with previous 2-D claims of Rayleigh-Taylor suppression \cite{perkins2017,srinivasan2013}. The magnetic field at this time is over $10^4 T$ in strength, with an estimated minimum unstable perturbation wavelength (from equation \ref{eq:lambda}) of $80\mu m$, which is much greater than the actual wavelength of $25\mu m$. Note that the plasma $\beta$ at this time is everywhere over 10, which would suggest that the Lorentz force is only of secondary importance; this highlights that the plasma $\beta$ is misleading when dealing with magnetic tension effects.
	
	While the perturbation is almost completely suppressed along the magnetic field direction, the total bubble-spike height remains similar to the unmagnetized case. In the $\underline{y}$ direction the spike is able to push between magnetic field lines. In other words, large amplitudes in $\underline{y}$ do not result in the bending of magnetic field lines, and there is no tension force to suppress growth.
	
	Figure \ref{fig:single_temporal} shows the temporal dependence of the spike-bubble height in $\underline{x}$ and $\underline{y}$ for a perturbation of wavelength $\lambda = 50 \mu m$. The height is measured using the $1keV$ electron temperature contour. In the $\underline{y}$ direction (figure \ref{fig:single_temporaly}) the spike is affected by classical growth and ablative stabilization only. At later times the thermal ablative stabilization becomes more dominant, as the plasma temperature increases. The applied field suppresses the heat-flow that drives ablation, resulting in a spike-bubble depth that for the 50T case is $60 \%$ larger than the unmagnetized spike at 2.1ns. At this time the Hall Parameter near the spike tip is approximately 4, which suppressed the heat-flow by more than 95\% ($\kappa_{\bot}/\kappa_{\parallel}<0.05$).
	
	%Similar plots can be produced by computing the line-integrated density $\rho R$ variation, but the metric for spike amplitude ($\Delta \rho R$) is less intuitive than a height. 

	In the  $\underline{x}$ direction (figure \ref{fig:single_temporalx}) the spike is also affected by magnetic tension. The applied field both acts to decrease the thermal ablative stabilization and increase the magnetic tension stabilization. For the 5T case these contributions roughly cancel, giving a spike height that is similar to the unmagnetized simulation at all times. For 50T there is significant stabilization even at early times, resulting in anisotropic perturbation growth; the ratio of the height in $\underline{y}$ to the height in $\underline{x}$ is $h_y/h_x > 15 $ by $2.1ns$.
	
	For the first time, resistive diffusion of the magnetic field is found to play an important role in mitigating the effectiveness of magnetic stabilization. Figure \ref{fig:single_temporalx} includes a plot of the $30T$ spike-bubble height when the magnetic diffusivity has been artificially reduced by a factor of 10 in the simulations. The diffusion allows for the perturbations to push through the field lines, and increases in importance with time. With resistive diffusion reduced, the peak spike height along the field lines is smaller and occurs approximately 0.15ns earlier in time, with the perturbation suppressed thereafter.

	To understand how magnetic tension and magnetized ablative stabilization vary for different length-scale perturbations, 3-D single-mode simulations have been completed over the range $\lambda=25\mu m, 100\mu m$. The resultant bubble-spike heights along $\underline{x}$ and $\underline{y}$ are plotted in figure \ref{fig:single_wavelength} at $1.9ns$. Both magnetic tension and ablative stabilization become most important at shorter wavelengths. In $\underline{y}$ the magnetic tension does not affect the perturbation growth; therefore, the applied field increases the bubble-spike height. In $\underline{x}$ the effect of magnetic tension dominates over the reduced ablative stabilization for all wavelengths with 30T and 50T applied fields. 
	
	Also plotted in figure \ref{fig:single_wavelengthx} is the 30T bubble-spike height in $\underline{x}$ with the resistive diffusion reduced by a factor of 10. A simple scaling has the diffusion time-scale behaving as $t_{diff} \sim \eta/\lambda L_m$, where $L_m$ is the density length-scale. This is consistent with the 3-D simulations, with the shortest wavelengths proportionately greatest affected by the diffusion. For $\lambda=40\mu m$ the spike-bubble height increases by $10 \mu m$ due to diffusion, while for $\lambda=100\mu m$ the impact of diffusion changes the perturbation amplitude by $2 \mu m$.
	
	%. Using frozen-in flow combined with a diffusion factor gives  $|\underline{B}|= B_0 (1- t/t_{diff})\rho/\rho_0$. Combined with equation \ref{eq:gammax} it is clear that the diffusion impacts the 

\begin{figure}
	\centering
	\includegraphics[scale=0.8]{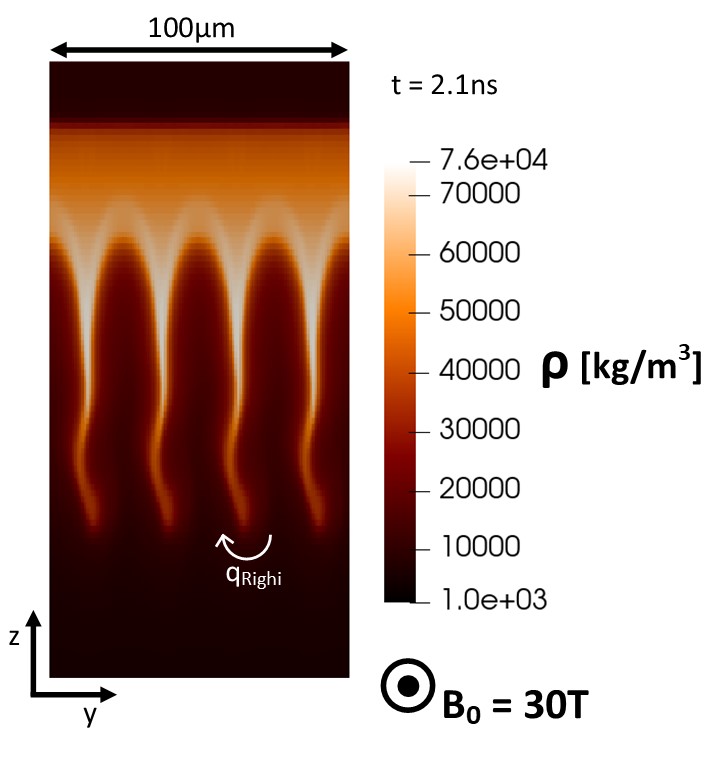}\caption{ \label{fig:xMHD} 2-D slice of mass density of a single-mode perturbation at 2.1ns with a 30T magnetic field applied out of the page. The Righi-Leduc heat-flow vector is shown, which acts to break the perturbation symmetry. }
	\end{figure}

	So far the discussion of magnetized electron heat-flow has been restricted to the suppression of thermal conduction perpendicular to field lines. Here it is also noted that the Righi-Leduc heat-flow (the $\kappa_{\wedge}$ term in equation \ref{eq:heatflow}) modifies perturbation growth, particularly for short wavelengths. Figure \ref{fig:xMHD} shows a 2-D $\underline{y}-\underline{z}$ slice through a simulation with perturbation wavelength of $25\mu$m at t=2.1ns. The applied field is 30T out of the page. The orientation of Righi-Leduc heat-flow is depicted, which breaks the left-right symmetry of the spike. Qualitatively, electrons that stream along $\underline{z}$ into the spike are deflected left by the magnetic field \cite{braginskii1965}. Righi-Leduc results in greater ablation on the left side of the spike than the right, which introduces a higher mode. For the cases simulated, the 30T applied field gives the greatest effect, with 5T resulting in too low magnetizations and 50T then suppressing the Righi-Leduc coefficient. 
	
	Biermann Battery magnetic fields have been included in all of the simulations results in this paper. The self-generated fields do moderately change the perturbation propagation. However, the process is detailed and will be the subject of a future publication. It suffices to say here that the impact of Biermann Battery is of secondary importance to the applied magnetic field and does not significantly change in significance depending on the applied field strength.

	\section{Multi-Mode \label{sec:multimode}}
	
	\begin{figure}
		\centering
		\includegraphics[width=0.5\textwidth]{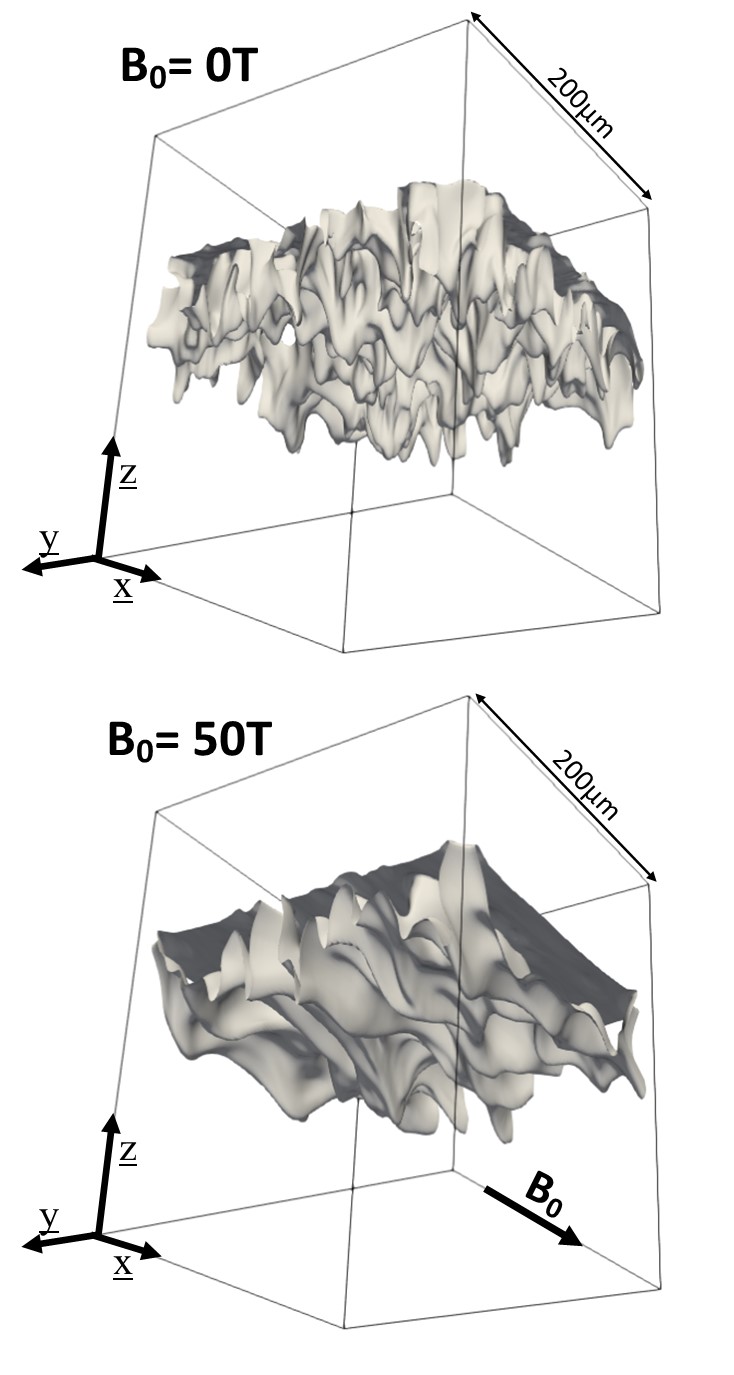}\caption{ \label{fig:multi_contours}  Density contours ($6 \times 10^{4} kg/m^3$) at 1.9ns for simulations perturbed with a multitude of modes. Left: no magnetic field applied. Right: 50T magnetic field applied along $\underline{x}$. These simulations use maximum initial perturbation sizes of $\epsilon = 0.5\mu$m.}
	\end{figure}

	\begin{figure}
		\centering
		\begin{subfigure}[b]{0.5\textwidth}
			\centering
			\includegraphics[width=1.\textwidth]{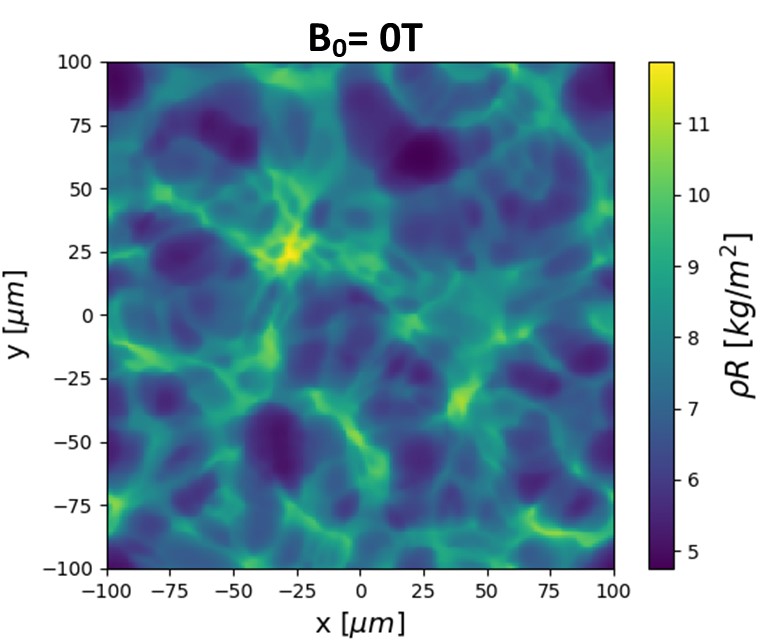}
			\caption{}
			\label{fig:multi_rhoR_0T}
		\end{subfigure}
		\begin{subfigure}[b]{0.5\textwidth}
			\centering
			\includegraphics[width=1.\textwidth]{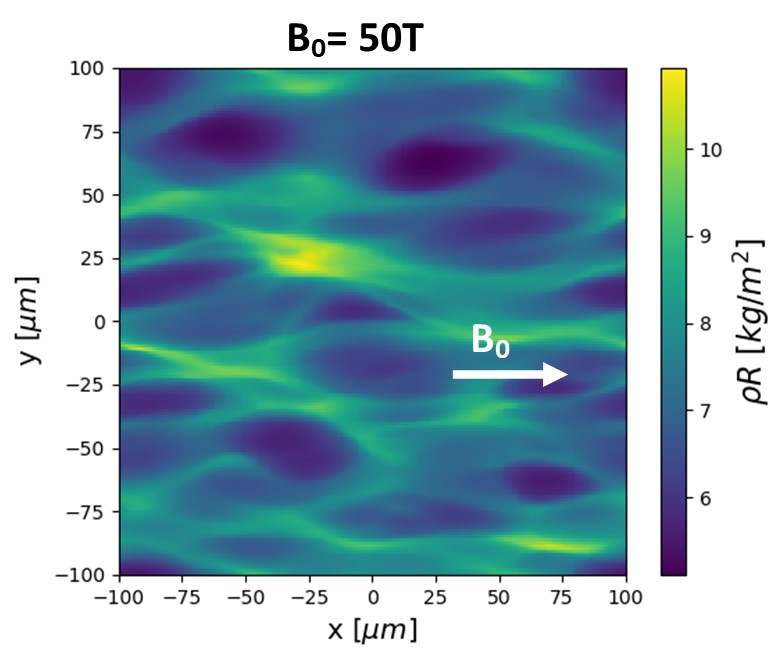}
			\caption{}
			\label{fig:multi_rhoR_50T}
		\end{subfigure}
		
		\caption{Mass density of a multi-mode simulations integrated along $\underline{z}$ to get the areal density with an applied field of a: 0T ; b: 50T. The initial perturbation size is set using $\epsilon=0.5\mu m$. The data is plotted here at 1.6ns. The magnetic field (applied from left to right) causes perturbation striations to grow.} 
		\label{fig:multi_rhoR}
	\end{figure}
	
	Multi-mode simulations use a domain with $200\mu$m extent in both $\underline{x}$ and $\underline{y}$. The perturbation is initialized using 500 individual modulations, each with an amplitude randomly chosen between $\pm \epsilon$. The wavelength is chosen randomly in the range $4 \mu$m, $200 \mu$m and the central location is chosen randomly in $\underline{x}$ and $\underline{y}$. Only one full wavelength of each cosine perturbation is applied, leaving the rest of the domain unperturbed. Both the initialization and evolution use periodic boundary conditions, such that the behavior at the domain edge is not affected by the boundary. $1/2 \mu $m resolution is used in $\underline{x}$ and $\underline{y}$; $1\mu$m resolution is used in $\underline{z}$. The simulations have been executed for multiple maximum perturbation sizes, $\epsilon = 0.2\mu$m, $0.5 \mu$m, $1.0\mu$m, $4.0\mu$m, with the random number generator staying the same between runs in order to keep the same distribution of modes.
	
	Density contours of $\rho = 6 \times 10^{14} kg/m^3$ are compared in figure \ref{fig:multi_contours} at 1.9ns for 0T and 50T with initial perturbation size $\epsilon = 0.5\mu m$. Similar to the single-mode case, the magnetic field suppresses perturbation growth in $\underline{x}$, forming striations with only small amplitude modulations. In $\underline{y}$, however, the bubble-spike amplitudes are similar to the unmagnetized case.
	
	To further show the developed striations, the density is integrated along $\underline{z}$ at 1.6ns, giving areal density as a function of $\underline{x}$ and $\underline{y}$. Figure \ref{fig:multi_rhoR} compares this result with and without the 50T applied field. When no magnetic field is imposed (figure \ref{fig:multi_rhoR_0T}) the perturbation growth has no preferential orientation. For a 50T applied field (figure \ref{fig:multi_rhoR_50T}), the striations clearly form, oriented along the applied field direction. 
	
\begin{figure}
		\centering
		\includegraphics[scale=0.55]{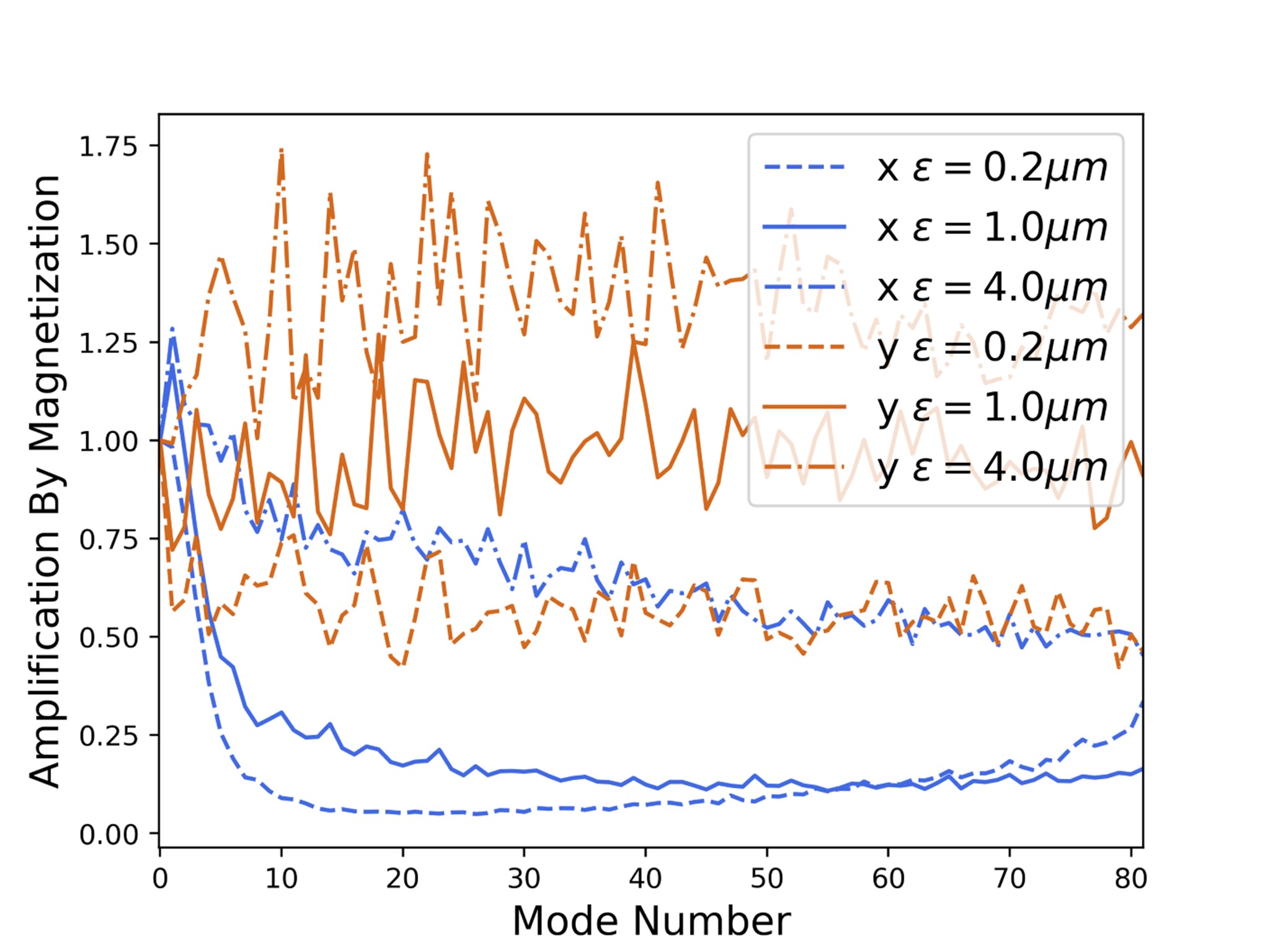}\caption{ \label{fig:multi_FFT}  Amplification of  $\rho R$ variation due to magnetic field as a function of mode number at $1.6ns$ for a 50T applied field along $\underline{x}$. Perturbations along $\underline{x}$ and $\underline{y}$ are shown separately, with modulations in $\underline{x}$ almost completely suppressed, while perturbations in $\underline{y}$ are only moderately hindered by the magnetic field.}
	\end{figure}

	This behavior can be quantified by taking the Fourier transform of this areal density along $\underline{x}$ and $\underline{y}$. These are plotted normalized to the 0T case in figure \ref{fig:multi_FFT} at 1.6ns for different initialized perturbation amplitudes. The suppression of perturbations along the magnetic field lines is severe, particularly for modes greater than 5. For the largest applied perturbation size, $\epsilon = 4.0\mu$m, the suppression is much less significant, only approximately 50\% lower than the unmagnetized case.
	
	Perpendicular to the magnetic field ($\underline{y}$) the impact of the magnetic field is also found to be dependent on the initial perturbation size. For $\epsilon=4\mu$m the reduced ablative stabilization by electron magnetization allows for the instability to grow larger, by around 30-40\% at 1.6ns; this result is consistent with the findings in section \ref{sec:singlemode} for single-mode perturbations. 
	
	As the initial perturbation size is lowered, however, the spike height in the $\underline{y}$ direction is lowered by magnetization. For $\epsilon=0.2\mu$m the suppression is up to 50\% by a 50T magnetic field.  A strong applied field results in the instability growing in only 2 dimensions (although not the two dimensions simulated by 2-D simulations \cite{perkins2017,srinivasan2013}). It has long been known that the Rayleigh-Taylor instability grows slower in 2-D rather than 3-D \cite{anuchina2004,zhou2017,zhou2019}. A factor of around 2-3 has been observed in previous non-linear simulations comparing dimensionality \cite{anuchina2004}. Therefore, the suppression of spike-bubble height in $\underline{x}$ can have a moderate and indirect impact on the spike height in $\underline{y}$. However, for larger amplitude perturbations (as seen in figure \ref{fig:multi_FFT}), the changes in thermal ablative stabilization dominate. 
	
	\section{Conclusions}
	
	In summary, the magnetized ablative-Rayleigh-Taylor instability has been simulated in 3-D for the first time. Previous 2-D simulations over-predicted the impact of magnetic tension stabilization by ignoring the most unstable dimension \cite{perkins2017,perkins2013,srinivasan2013,PhysRevE.104.L023201}. For spike modes perpendicular to the magnetic field ($\underline{\hat{k}}\cdot\underline{\hat{b}} = 0$) growth is, at best, reduced from a 3-D growth rate to 2-D, which is approximately a factor of 2 smaller. For larger perturbation amplitudes, where the thermal ablative stabilization is more important, the peak bubble-spike height can be increased by magnetization. 
	
	Short wavelengths are shown to be most affected by magnetization. The magnetic tension stabilizes along the field direction, but reduces ablative stabilization perpendicular, resulting in anisotropies in spike-bubble heights $h_y/h_x > 30$ in the most extreme cases simulated.
	
	Resistive diffusion has also been found to modify the magnetic tension stabilization. This is particularly true for short wavelengths and long time-scale evolution, where the resistivity allows the spike to push through the magnetic field. Righi-Leduc heat-flow has also been shown to induce higher modes onto spikes that have electron magnetizations $\omega_e \tau_e \approx 1$.
	
	These results have significant implications for magnetized spherical and cylindrical implosions on laser and pulsed-power ICF facilities \cite{perkins2017,chang2011,slutz2010,davies2017}. While magnetic fields are still anticipated to increase fusion performance, the impact of fields on perturbation growth are shown here to be dependent on dominant perturbation mode number, amplitude and applied field strength. For spherical implosions dominated by axi-symmetric asymmetries (such as low convergence indirect-drive implosions on the National Ignition Facility) 2-D simulations including the magnetic tension stabilization may be sufficient; P2 and P4 low-mode asymmetries as well as the tent scar fall into this category \cite{clark2016}. However, assessing the impact of a magnetic field on surface roughness, beam imprint or fill-tube asymmetries, will require 3-D calculations. In cases where this is intractable, 2-D calculations should be conducted both with and without the Lorentz force included, in order to assess the extreme cases. 
	
	The impact of an applied magnetic field on the mix induced by a fill-tube is of particular interest to the ICF community; as the fill-tube is attached to the capsule waist it will propagate perpendicular to an external magnetic field. The induced short wavelength features that are responsible for the mix of high-Z material into the hot-spot are expected to be highly susceptible to stabilization along the magnetic field lines. Perpendicular to the field, however, the impact of the magnetic field is less clear. This work motivates the execution of computationally-intensive 3-D simulations of a magnetized fill-tube in order to shed light on the potential gains of magnetizing ICF targets.
	
	\section*{Acknowledgements}
		This work was performed under the auspices of the U.S. Department of Energy by Lawrence Livermore National Laboratory under Contract DE-AC52-07NA27344 and by the LLNL-LDRD program under Project Number 20-SI-002.
		
		This document was prepared as an account of work sponsored by an agency of the United States government. Neither the United States government nor Lawrence Livermore National Security, LLC, nor any of their employees makes any warranty, expressed or implied, or assumes any legal liability or responsibility for the accuracy, completeness, or usefulness of any information, apparatus, product, or process disclosed, or represents that its use would not infringe privately owned rights. Reference herein to any specific commercial product, process, or service by trade name, trademark, manufacturer, or otherwise does not necessarily constitute or imply its endorsement, recommendation, or favoring by the United States government or Lawrence Livermore National Security, LLC. The views and opinions of authors expressed herein do not necessarily state or reflect those of the United States government or Lawrence Livermore National Security, LLC, and shall not be used for advertising or product endorsement purposes.
		
	\section*{Data Availability}
	The data that support the findings of this study are available from the corresponding author upon reasonable request.
		
	\section*{References}
	
	\ifdefined\DeclarePrefChars\DeclarePrefChars{'’-}\else\fi

\end{document}